\documentclass{LT23auth}
\usepackage{graphicx}
\usepackage{bm}      
\usepackage{amssymb}

\begin{document}

\begin{frontmatter}

\title{
Zero-energy edge states and their origin 
in particle-hole symmetric systems:
symmetry and topology
}

\author[address1]{Shinsei Ryu\thanksref{thank1}},
\author[address2]{Yasuhiro Hatsugai},

\address[address1]{
Dept. of Applied Physics, Univ. of Tokyo, 7-3-1 Hongo,
Bunkyo-ku, Tokyo, Japan
}

\address[address2]{
Dept. of Applied Physics, Univ. of Tokyo, 7-3-1 Hongo,
Bunkyo-ku, Tokyo, Japan and \\
PRESTO, JST, Saitama 332-0012, Japan
}

\thanks[thank1]{Corresponding author. 
 E-mail: ryuu@pothos.t.u-tokyo.ac.jp}

\begin{abstract}
We propose a criterion to determine
the existence of zero-energy edge states for a class of
particle-hole symmetric systems.
A loop is assigned for each system,
and its topology and a symmetry play an essential role.
Applications to $d$-wave superconductors are demonstrated.
\end{abstract}

\end{frontmatter}

\paragraph*{Introduction}
When a quantum system is terminated to a finite size,
it may support a state localized at the boundaries.
The appearance of such states is a hallmark of 
a phase degree of freedom 
specific to quantum mechanical systems,
and has been investigated in the context of, say,
quantum Hall or Haldane spin systems.
In addition to these gapped cases,
edge states in gapless systems have attracted
a great deal of interest recently.
For example,
zero-energy edge states (ZES)
at a $(110)$ surface in $d$-wave superconductors (SC)
have been observed via a tunneling spectroscopy.
\cite{hu,tanaka,iguchi}
Graphite ribbons are also known to support
ZES, which lead to several physical consequences 
such as spin polarization near the boundaries. \cite{fujita}
In this article, we present a criterion to tell
whether a certain system supports ZES.
Our criterion is built on a symmetry and topology.
Applications to ZES in $d$-wave SC is demonstrated.
We also discuss an instability caused by ZES for $d$-wave SC:
the emergence of the time-reversal symmetry breaking 
superconducting order parameters near (110) surfaces.
\cite{matsumoto}

\paragraph*{Criterion for ZES}
A class of systems that we are concerned with
is described by the following single-particle hamiltonian:
\begin{eqnarray}
{\mathcal H}=
\sum_{x,x^{\prime}}
\boldsymbol{c}_{x}^{\dagger}h_{x,x^{\prime}}
\boldsymbol{c}_{x^{\prime}},
&&
h_{x,x^{\prime}}=
\left[
\begin{array}{cc}
t_{x,x^{\prime}} & \Delta_{x,x^{\prime}} \\
\Delta^{\prime}_{x,x^{\prime}} & -t_{x,x^{\prime}}
\end{array}
\right]=h_{x^{\prime},x}^{\dagger},
\nonumber 
\end{eqnarray}
where 
$
t_{x,x^{\prime}},\Delta_{x,x^{\prime}},\Delta^{\prime}_{x,x^{\prime}}
\in \mathbb{C}
$,
and
$\boldsymbol{c}^{\dagger}_{x}=(c^{\dagger}_{x\uparrow},c_{x\downarrow}^{\vphantom{\dagger}})$
denotes electron creation/annihilation operators at site $x$.
The system is defined on a 1D lattice,
with its total number of the lattice sites being $N_{x}$,
and $x=1,\cdots , N_{x}$.

Our criterion for ZES is stated in terms of the bulk properties
of the system and the shapes of edges.
The bulk property of the system is easily captured 
by adopting the periodic boundary condition.
With the periodic boundary condition,
the above hamiltonian can be transformed to
\begin{eqnarray}
{\mathcal H}^{\mathrm{bulk}}&=&\sum_{k}
\boldsymbol{c}_{k}^{\dagger}\,
\boldsymbol{R}(k)\cdot \mathcal{\sigma}
\,
\boldsymbol{c}_{k},
\nonumber 
\end{eqnarray}
where
$k\in (-\pi,\pi]=S^{1}$ is the crystal momentum,
and $\sigma_{X,Y,Z}$ the Pauli matrices.
All bulk properties are encoded in the Fourier-transformed matrix element 
$\boldsymbol{R}(k) \in \mathbb{R}^{3}$,
from which
we can identify a loop $\ell:
k\in S^{1}\rightarrow \boldsymbol{R}(k)\in \mathbb{R}^{3}$
for each 1D Hamiltonian ${\mathcal H}^{\mathrm{bulk}}$.
A system with a certain type of edges is then generated
by truncating a bulk Hamiltonian ${\mathcal H}^{\mathrm{bulk}}$.
We refer a prescription for creating edges as $e$.
Generally, $e$ represents an impurity potential at an edge, 
coexistence of different order parameters near boundaries
in superconducting systems, etc.
We write a system characterized by $\ell$ and $e$ as
${\mathcal H}^{\mathrm{edge}}[\ell,e]$.
Then, we ask if
${\mathcal H}^{\mathrm{edge}}[\ell,e]$ supports
ZES localized at either end of the sample
for given $\ell$ and $e$.

Our criterion to tell the existence of ZES is 
summarized as follows\cite{ryu}:
\begin{description}
\item[(A)]
The loop $\ell$ is on a plane cutting the origin ${\mathcal O}$
of the three-dimensional $\boldsymbol{R}$-space.
For loops that satisfy this condition,
we can find an operator $\Gamma$ which anticommutes with
${\mathcal H}^{\mathrm{bulk}}[\ell]$.
We call this property as chiral symmetry.
$\Gamma$ is equivalent to $\mathbf{1}_{N_{x}}\otimes \sigma_{Z}$ 
upto an arbitrary SU$(2)$ transformation,
where ${\mathbf 1}_{N_{x}}$ acts on a site index,
while $\sigma_{Z}$ on a spin index.
\item[(B)]
$\ell$ can be continuously deformed to $\ell_{c}$
without crossing ${\mathcal O}$,
where $\ell_{c}$ is a unit circle on the plane.
$(\ell \sim \ell_{c})$
\item[(C)]
$e$ respects the chiral symmetry.
That is, ${\mathcal H}^{\mathrm{edge}}[\ell,e]$ anticommutes
with $\Gamma$.
\end{description}
If the conditions (A)-(C) are satisfied,
there exists at least a pair of ZES,
one of which localized at the right edge and 
the other at the left edge.

For 2D or higher-dimensional systems with edges, 
we first Fourier transform 
along directions parallel to the edge,
to get a family of 1D Hamiltonians
parametrized by the wave number along the edge.
Then, the present results are applicable for each 1D Hamiltonian.

\begin{figure}[btp]
\begin{center}\leavevmode
\includegraphics[width=\linewidth]{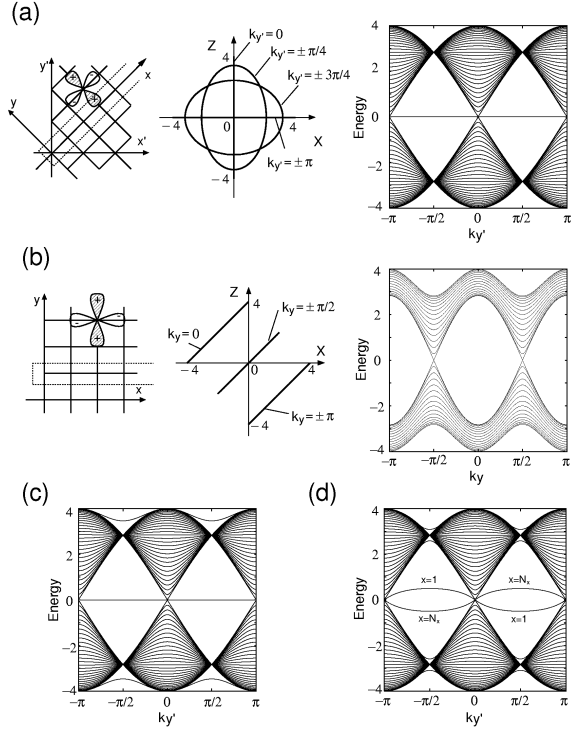}
\caption{ 
Loops in the $\boldsymbol{R}$-space and the corresponding 
energy spectrum for $d_{\mathrm{x}^{2}-\mathrm{y}^{2}}$ SC with
(a) $(110)$ and (b) $(100)$ surfaces.
Energy spectra with
introduction of (c) $s$ and (d) $\mathrm{i}s$ order parameter
near a $(110)$ surface.
An edge mode localized at the site $1,N_{x}$ 
is indicated by $x=1,N_{x}$.
}
\label{LT}
\end{center}
\end{figure}

As an application of the present results,
let us discuss 2D $d_{\mathrm{x}^{2}-\mathrm{y}^{2}}$-wave SC with surfaces.
In Fig.(\ref{LT}a) and (\ref{LT}b),
a family of loops in $\boldsymbol{R}$-space and the energy spectrum is shown 
for  $d_{\mathrm{x}^{2}-\mathrm{y}^{2}}$-wave SC with 
(a) $(110)$ and (b)$(100)$ surfaces.
For the $(110)$ case,
loops are an ellipsis on the $XZ$-plane enclosing ${\mathcal O}$
except at $k_{y^{\prime}}=\pm\pi,0$ 
($k_{y^{\prime}}$ is the wave number along the edges).
Thus,the above criterion tells us there are ZES for the $(110)$ case.
On the other hand,
we do not expect ZES for the $(100)$ case since 
loops collapse into a line segment.
We have verified numerically this 
prediction in Fig. (\ref{LT}b).

\paragraph*{Peierls-like instability and the chiral symmetry breaking}
Since edge states with different $k_{y^{\prime}}$ are 
all degenerate at $E=0$,
they are expected to cause a Peierls-like instability.
In presence of interactions,
parameters in a single particle Hamiltonian
$t,\Delta,\Delta^{\prime}$ near the edges
might be effectively modified
in order to lift the degeneracy,
and thereby lower the ground state energy.
However, since these ZES are stable 
to perturbations which respect the chiral symmetry
(Statement (C)),
such modifications should be accompanied 
with the breaking of the chiral symmetry
near the boundaries.
The emergence of, say, $\mathrm{i}s$ or $\mathrm{i}d_{\mathrm{xy}}$ components 
near the boundary can breaks the chiral symmetry
to lift the degeneracy of edge modes.
On the other hand, a purely real order parameter cannot do it.
Indeed, it has been revealed via a quasi-classical study that
coexistence of $\mathrm{i}s$- or 
$\mathrm{i}d_{\mathrm{xy}}$-wave order parameter
with $d_{\mathrm{x}^{2}-\mathrm{y}^{2}}$-wave 
near the surface is possible for the $(110)$ surface.
\cite{matsumoto}

This is explicitly demonstrated in Fig. (\ref{LT}c) and (\ref{LT}d).
With the introduction of $\mathrm{i}s$ order parameter
near a $(110)$ surface in $d_{\mathrm{x}^{2}-\mathrm{y}^{2}}$ SC,
the flat band formed by edge states develops a finite dispersion.
On the other hand, introduction of $s$ order parameter
does not break the chiral symmetry, and hence
cannot lift the degeneracy.

%
%

\end{document}